# AN ANALYSIS OF KEY FACTORS FOR THE SUCCESS OF THE COMMUNAL MANAGEMENT OF KNOWLEDGE

Isabelle Bourdon

Professeur en Système d'Information

Université Montpellier II

Montpellier

France

isabelle.bourdon@polytech.univ-montp2.fr

Chris Kimble

Lecturer

University of York

York

UK

kimble@cs.york.ac.uk

*Abstract*

*This paper explores the links between Knowledge Management and new community-based models of the organization from both a theoretical and an empirical perspective. From a theoretical standpoint, we look at Communities of Practice (CoPs) and Knowledge Management (KM) and explore the links between the two as they relate to the use of information systems to manage knowledge. We begin by reviewing technologically supported approaches to KM and introduce the idea of "Systèmes d'Aide à la Gestion des Connaissances" SAGC (Systems to aid the Management of Knowledge). Following this we examine the contribution that communal structures such as CoPs can make to intraorganizational KM and highlight some of 'success factors' for this approach to KM that are found in the literature. From an empirical standpoint, we present the results of a survey involving the Chief Knowledge Officers (CKOs) of twelve large French businesses; the objective of this study was to identify the factors that might influence the success of such approaches. The survey was analysed using thematic content analysis and the results are presented here with some short illustrative quotes from the CKOs. Finally, the paper concludes with some brief reflections on what can be learnt from looking at this problem from these two perspectives.*

# 1 Introduction

The focus of this paper is principally on intraorganizational Knowledge Management (KM) and the contribution that communal structures such as Communities of Practice (CoPs) can make to this. The concept of a firm based on knowledge places organizational knowledge as the main source of competitive advantage for that organization. Such firms are thought of as entities that create value through managing their knowledge. Developments in Information Technology, coupled with increasing awareness of the importance of organizational knowledge, have lead to the development of a variety of information systems that attempt to manage this knowledge. If we accept this view of the firm, then the identification of factors that will encourage the exchange of knowledge become a top priority; it is this question that constitutes the heart of our research.

Notwithstanding the drive toward using information systems to manage knowledge, there is also a recognition that most organizational knowledge remains firmly anchored in individuals (Nonaka and Konno 1998), and consequently, the KM process is highly dependent on the behaviour of the employees within an organization. The resistance of certain groups or individuals frequently impedes the intraorganizational knowledge sharing (Ciborra and Patriota 1998); this is why many organizations are now beginning to look at community-based models, which, it is believed, will encourage the sharing of knowledge (Davenport et al. 1998).

Communities of Practice (CoPs) appear to be one of the most favoured organizational forms to encourage the sharing of organizational knowledge. Argote et al. (Argote et al. 2003) identified a set of emergent issues for the future of research on KM. They highlight the "*importance of social relations in understanding knowledge creation, retention and transfer*" and "*the fit between properties of knowledge and properties of relationships in a social system*" (Argote et al. 2003). They point to the need to shift our interest from single to multiple relations when dealing with the KM process. Such indications suggest that we should concentrate our research efforts on achieving a more complete understanding of key features of community-based organizations.

This paper reports on a study of twelve large French companies undertaken to identify the factors influencing the success of community based approaches to intraorganizational KM. It was anticipated that the identification of such factors would allow organizations to gain a better understanding of the reasons why their attempts to share knowledge were successful or unsuccessful.

## 2  Towards the Communal Management of Intraorganizational Knowledge

### 2.1   The Processes of Intraorganizational KM

For Reix (Reix 1995), *"the knowledge held by an enterprise is a major element of its competitive advantage"*. There are many definitions of knowledge (Hildreth and Kimble 2002) however, most are specific to the context in which they are used. From the KM perspective, Davenport and Prusak (2000) observe:

> *"Knowledge is a fluid mix of framed experiences, values, contextual information and expert insight that provides a framework for evaluating and incorporating new experiences and information. It originates and is applied in the minds of knowers. In organizations, it often becomes embedded not only in documents or repositories but also in organizational routines, processes, practices and norms". (Davenport and Prusak 1998)*

While Schreiber *et al*. (1999), from a Knowledge Engineering perspective, define knowledge as:

> *"... the whole body of data and information that people bring to bear to practical use in action, in order to carry out tasks and create new information. Knowledge adds two distinct aspects: first a sense of purpose, since knowledge is the 'intellectual machinery' used to achieve a goal; second, a generative capability, because one of the major functions of knowledge is to produce new information". (Schreiber et al. 1999)*

Although both have a slightly different definition for knowledge, their focus is the same: knowledge is a resource that needs to be managed. Information and Communication Technologies (ICTs) are used to help sustain the processes of KM within organizations (Alavi and Leidner 2001). Tools known as Knowledge

Management Systems (KMS) are increasingly seen as a way to take advantage of opportunities for exchange and knowledge sharing (Knowings 2002).

Two main models of KMS have been identified in the Information Systems literature: the index model and the social network model.

- The index or integrative model corresponds to a codification approach to KM (Hansen et al. 1999). This approach focuses on the codification and the storage of knowledge in order to facilitate its reuse through access to the codified data. The key technological component of this approach is the electronic knowledge index (Grover and Davenport 2001).

- The social network or interactive model corresponds to a customization approach to KM (Hansen et al. 1999). This approach focuses on the links between individuals for the exchange of knowledge. A key technological component of this approach are knowledge maps, that assure the localization of knowledge and tools that allow individuals to interact (Hildreth et al. 1999).

In order to combine the advantages of the index model with the network model, organizations have moved from global KMS to systems of a more communal type, which are local and centred on particular interests. For example a number of French enterprises (e.g. Arcelor, Bouygues Telecom, Bureau Veritas, CGEY, EDF-GDF, Gemplus, La Poste, Renault, Rhodia, Sanofi-Synthelabo, Schneider Electric, ST microelectronics, Total and Valeo), have introduced communal KMS as part of their KM strategy.

In organizations of this type, groups can materialize out of virtual communities through the use of shared technological networks through Information Systems known as *'Systèmes d'Aide à la Gestion des Connaissances'* (SAGC) - literally "Systems to aid the Management of Knowledge" - (Bourdon et al. 2003; Bourdon et al. 2004). These developments are based on the grouping of individuals in communal structures that are favourable to collaboration and to the sharing of internal knowledge.

## 2.2 The Contributions of Communal Structures to the Management of Knowledge

CoPs are groups of persons that share an interest, a subject or a common problem and deepen their knowledge of it through participation in its practice. This form of organization is used increasingly as a basis to develop the sharing of internal knowledge and for the construction of computer based tools for KM (Abdullah et al. 2006). The literature on the KM puts the accent on the fact that the CoPs can encourage the management of knowledge and can constitute the organizational structures needed to support KM. For example, a recent survey by IBM Global Services and Knowings of 200 leaders of enterprises in France, identified CoPs as the 'base cells of the knowledge sharing' (Knowings 2002).

Lave and Wenger (Lave and Wenger 1991) first introduced the notion of CoPs defining them as:

> *"... a set of relations among persons, activity, and world, over time and in relation with other tangential and overlapping Communities of Practice".*
> *(Lave and Wenger 1991, p98)*

Later, Wenger (Wenger 1998b) redefined CoPs as groups of people that share a topic of interest, common problems or a passion. CoPs are seen as groups that form spontaneously and whose members work together to share knowledge:

> *"Communities of Practice are important to the functioning of any organization, but they become crucial to those that recognize knowledge as a key asset ... Knowledge is created, shared, organized, revised, and passed on within and among these communities." (Wenger 1998a)*

Furthermore, organizations themselves are seen as being 'constellations' of CoPs:

> *"From this perspective, an effective organization comprises a constellation of interconnected Communities of Practice, each dealing with specific aspects of the company's competency - from the peculiarities of a long-standing client, to manufacturing safety, to esoteric technical inventions." (Wenger 1998b, p 127)*

In a departure from his earlier collaboration with Lave (Lave and Wenger 1991), Wenger (1998b) argues that CoPs can be characterised in terms of dualities:

> *"... a single conceptual unit that is formed by two inseparable and mutually constitutive elements whose inherent tensions and*

> *complementarity give the concept richness and dynamism"* *(Wenger 1998b, p 66)*

He identifies four such dualities: participation-reification, designed-emergent, identification-negotiability and local-global. Because of its obvious links to index models of KM, the participation-reification duality has been the focus of particular interest in this field.

Wenger also provides (1998b, pp 72 - 73) a concise definition for a CoP consisting of just three interrelated terms: 'joint enterprise', 'mutual engagement' and 'shared repertoire'. Members interact with one another and, in doing so, establish social norms and build relationships; this is termed mutual engagement. Secondly, through their interactions, they create an understanding of the shared interests that bind them together; this is termed the joint enterprise. Finally, over time, they produce a set of communal resources, termed their shared repertoire, which they use in the pursuit of their joint enterprise. This shared repertoire can include both symbolic and literal meanings such as, symbols, rituals and language as well as physical artefacts such as documents, files or, in this context, a SAGC.

## 2.3   The Potential Contribution of CoPs to the Management of Intra-organisational Knowledge

CoPs are seen as organizational structures that drive the individuals, through their common interest, to share their knowledge and expertise (Davenport et al. 2001; Grover and Davenport 2001). For example, Hildreth and Kimble (Hildreth and Kimble 2002; Kimble and Hildreth 2005) highlight the role CoPs in knowledge sharing in organisations; in particular, the way that strong communal ties between individuals (Constant et al. 1996) and the existence of a shared social capital (Nahapiet and Ghoshal 1998), constitute favourable conditions for the sharing of knowledge. The argument is that it is through mechanisms such as these, that knowledge sharing can be better understood (Boland and Tenkasi 1995).

We have summarized key points from some of the main works that deal with the contributions of CoPs to the management of knowledge and knowledge sharing in Table 1 below.

| Authors | Foundations |
|---|---|
| (McLure and Faraj 2000) | In a CoP, the exchange of knowledge is motivated by a moral obligation or interest in the community rather than by personal interest. |
| (Brown and Duguid 1991) | The flows of knowledge are better through networks of individuals who share the same interests in their work. |
| (Jarvenpaa and Staples 2000) | When individuals are encouraged to share their knowledge within a CoP, the cultural barriers to knowledge transfer weaken. |
| (O'Dell and Grayson 1998) | The lack of contact, relationships and common outlook between individuals is a manifest barrier to the transfer of the knowledge. |
| (Pan and Leidner 2002) | The importance of CoPs in the management of knowledge and the role that IT can play in sustaining knowledge sharing inside and between CoPs. |
| (Hildreth et al. 1999) (Hildreth and Kimble 2002) (Kimble and Hildreth 2005) | The role of CoPs in knowledge sharing: notably the role of shared artefacts, such as a SAGC, and the role of face-to-face communication. |
| (Vaast 2002) | The principal features of a CoP can be sustained by the use of an intranet. The feeling of belonging to a CoP is very important. |
| (Hall 2001) | The environments that encourage CoPs are also more favourable to the activities of sharing knowledge. |
| (Lefebvre et al. 2004) | The conditions for the spontaneous emergence of a CoP based on knowledge sharing can exist inside a research and development unit. |
| (Dyer and Nobeoka 2000) | The relationships within a CoP, characterized as 'mutual causality', are both a cause and a consequence of the process of learning and sharing of knowledge. |

**Table 1: The potential contributions of CoPs to KM**

Having established our theoretical foundations, we now wish to explore, in a more empirical manner, the factors that are perceived by the communities 'on the ground' to be favourable to the sharing of knowledge, particularly when using SAGCs.

## 3  An Exploratory Qualitative Survey

In order to identify the factors of CoPs perceived to be favourable to the sharing of knowledge via communal information systems, we analyzed the views of the people in charge of communal management system of knowledge. After describing the methods used in data gathering and analysis, we will present the main results of our analysis. A similar study using more quantitative approaches (Bourdon et al. 2004) has also been undertaken, although the results of that study do not feature here.

## 3.1 The Methodology

The study was carried out using semi-structured interviews on a set of Chief Knowledge Officers (CKOs) in large French companies. This instrument was used for its flexibility and the wealth of information it can generate (Miles and Huberman 1991). The survey was run in two stages, the first exploratory and the second based on an amended interview guide.

We were not overly directive in our face-to-face interviews with the CKOs. The interviews lasted from 30 minutes to 2.5 hours and were held in the officers' place of work. The first interview guide listed the main themes and sub-themes to discuss in the interview and was drafted beforehand. The second stage consisted of focused semi-directive interviews based on an amended interview guide, which enabled us to check and pinpoint the explanatory factors from the initial interview. Thirteen interviews were carried out in twelve companies.

| Company name | Job title of the interviewee | Length of interview |
|---|---|---|
| E | Chief Knowledge Officer | 1 hr |
| E | Chief Knowledge Officer | Attendance at a meeting |
| Er | Chief Knowledge Officer | 30 min |
| Er | Chief Knowledge Officer | 1 hr |
| G | Chief Information Officer | 1.5 hrs |
| R | Chief Knowledge Officer | 1 hr |
| S | Head of the SAGC office | 1 hr |
| S | Head of the SAGC office | 1 hr |
| S | Community leader | 1 hr |
| Sa | Chief Information Officer | 1 hr |
| T | Head of the intranet office | 1.5 hrs |
| T | Head of the intranet office | 1 hr |
| U | Chief Information Officer | 1.5 hrs |

**Table 2: survey sample**

For our data analysis and interpretation, we chose the thematic content analysis method, which is based on a system of themes and sub-themes (Berelson 1952). The premise of content analysis is that repetition of certain speech units (such as words, phrases, sentences or paragraphs) points to centres of interest and the opinions of the speakers. We defined our units of analysis as sentences, parts of sentences or groups of sentences, and then grouped them based on thematic content.

## 4 Results

We now present the key results of our qualitative survey into the factors that seem to influence the success of communal approaches KM. The survey highlights two types of factors present in community-based organizations that appear to encourage the sharing of knowledge: (1) Factors related to the characteristics of a CoP (2) Factors related to the organizational context. Due to constraints of space, we cannot give all the occurrences of the themes we identified during the interview; consequently we simply present a few key examples of 'units of meaning' accompanied by illustrative quotes from SAGC officers.

### 4.1 Factors related to the characteristics of a CoP

The thematic analysis of the interviews pointed to some underlying features of CoPs that were perceived to be important for knowledge sharing via SAGCs.

| |
|---|
| The pre-existence of a community |
| An understanding of the community |
| The formal structure of the community |
| The size of the community |
| The level of cooperation in the community |
| The vitality of the exchanges in the community |
| The presence of a shared repertoire within the community |
| The existence of clear positive benefits |
| The existence of a standard or indicator for the community |
| The quality of the exchanges |
| Trust within the community |
| The time allocated to community activities |
| The existence of rituals in the community |

**Table 3: Factors related to the characteristics of the CoPs**

**The pre-existence of a community.** It appeared that the presence of a community, or the perception of its pre-existence, was an element that the managers that we questioned considered a key factor in the success of this form of KM. For example:

> *"[…] in general when a network of people exists, it facilitates the technology [of the SAGC]; things go rather better […] then when one injects the technology […] when there are no pre-existent networks, it is true, the exercise is a lot more difficult."*

**An understanding of the community.** A deep understanding of the CoP itself was also seen as an important element. Understanding the focus of the community, the what, the how, its boundary, its composition and its goals. For example, an interviewee thinks:

> *"[…] one needs […] to identify those communities; to know what should exist […] after, [the community has bedded in] it needs to exist for the company, not locally in a bubble"*

**The formal structure of the community.** Setting up a formal structure for the underlying CoP in relation to the SAGC is also an important element in the sharing of knowledge. Thus, the apportionment of roles or functions dedicated to the community, the procedures and rules, the modalities of management and leadership are seen as essential factors. For example, the creation of the function of *animateur* or the leader of the community is seen as a key factor for the success of this approach:

> *"[every community] has one or two leaders […] who are nominees elected by the members of the community, and who belong to the community"*

**The size of the community.** The theme of the 'critical mass' of the community was considered crucial by the interviewees, for example:

> *"[…] the X community included 800 people and it didn't work at all, there was no activity, it had died. When one studied it, one found that there was too much diversity, it was varied too much, the X community was too large, therefore it divided in 13 different communities."*

**The level of cooperation in the community.** Another element is, of course, connected to the perceived level of cooperation. For example:
> *"[…] it is necessary to generate patterns of interactions and cooperation in communities where people are placed; but one cannot master all of these networks; on the contrary, one must act from within"*

**The vitality of the exchanges in the community.** This theme is also close the notion of the vitality of the exchanges within the community, for example:
> *"[…] what it is necessary to get [the CoP going] is a bandwagon effect, because, in a community of experts, there is […] 30% of the community that will go there, then, there are the others that are baffled because they don't go there."*

**The presence of a shared repertoire within the community.** The creation of a shared repertoire within the community is indispensable in terms of vision, common language, terminology, norms, principles, securities or beliefs. For example:
> *"[…] the essential first, it is to become aware that behind the obvious differences, there are the common goals, [which allow the] formalizing of knowledge and experiences"*

**The existence of clear positive benefits.** The perception of the benefits of exchange within the community is essential. Indeed, according those we questioned, individuals will be contributors only if they can see positive benefits. From an individual viewpoint, a manager states:
> *"[…] the members of the community go there because it gives them something for their own work."*

**The existence of a standard or indicator for the community.** Concerning organizational benefits, one manager stated:
> *"[…] it is necessary to be able to demonstrate the benefit within the group."*

The perception of the performance and results of the CoP in terms of a standard indicator is a critical characteristic. For example:

> *"To understand this success, we have accumulated stories, there have been 15 success stories across all communities, that say why it was a success, and in the end, [the community] was voted a success based on all the stories"*

**The quality of the exchanges.** The perception of the quality of the exchanges within the community is considered another factor that encourages sharing. So an interviewee indicates:

> *"In the system [...] what is difficult to measure, it is the relevance of the answer, it would be necessary to put the questioner's feedback in place such as - this answer helped me indeed, it save me so much time."*

**Trust within the community.** Trust within the CoP is perceived to be a factor encouraging the sharing of knowledge:

> *"[…] one is going to share, if one knows people, if one respects them, if one has trust in them and if it gives one pleasure [to share] with them in a protected space."*

**The time allocated to community activities.** Interviewees also felt that time restrictions were other elements that could influence individual participation to community. Therefore, the matter of the time allocated to community activities cropped up naturally when they talked about collaboration. For instance:

> *"[…] devoting time to building the community is no easy matter; time restrictions, which were not so common before, are now a major issue, [...]. Even when people are willing to do it, it's not necessarily their main work"*

**The existence of rituals in the community.** Finally, the setting-up of key events that are often cited as success factors, for example:

> *"[in the domain of the quality], the quality community exists, there is a big meeting twice a year with good grub for the 150 people in charge of the quality that lasts all day […] it is a festive event and people want to come and [...] after they met and communicated, there were some further exchanges."*

## 4.2 Factors related to the organizational context

Secondly, we present the main topics raised by the interviewees with regard to the organizational context.

| |
|---|
| The support of management |
| The provision of resources for the community |
| An organizational structure that facilitates collaboration |
| An organizational culture of sharing |
| Training in the use of supporting technologies |
| Appropriate systems of evaluation and incentives |
| The support of HRM |

**Table 4: Factors related to the characteristics of the organizational context**

Different interviewees insisted, repeatedly, that certain aspects of the organisation itself acted as inhibitory or facilitating factors for knowledge sharing via SAGCs. Strictly speaking, these factors do not appear to be specific to either CoPs or SAGCs; although all of the questions that were asked were asked in that context; it is interesting to note that many similar responses can be found in other studies of the use of information systems in organisations (e.g. Kimble and McLoughlin 1995)

**The support of management.** Content analysis of the interviews revealed that the influence of colleagues, managers and senior executives was a recurrent theme; concerning the influence of the management:

> *"[…] management has to be really involved in this approach, because locally it can have an effect on contributors in the field"*

Likewise

> *"[…] one condition for it to work, is that it must be a priority for the company, […] directors have to be visibly committed."*

**The provision of resources for the community.** The support for the community is also demonstrated by the provision of resources for the community:

> *"[…] in X the communities are alive, they are a little institutionalized, they have a minimum income, and are more or less developed"*

Resources however can also be human, for example, one interviewee commented that,

> *"[…] it worked, because people, experts on the subject, people with community recognition, guys of 55 reputed to be the best in the field, were assigned to the project to work on the aspects of building expertise"*

**An organizational structure that facilitates collaboration.** The structure of the organizations was considered by a number of managers we questioned to be a success factor e.g. the existence of a structure facilitating the collaboration rather than of competition between units. The organizational structure can enable or inhibit the sharing of knowledge. In particular, it affects the possibility for knowledge transfer across organizational units and/or hierarchical levels:

> *"[…] as long as company organization is vertical, with departments and divisions, even if you have matrices, as long as you stick to this logic, you create compartments that are responsible for obstructing outreach networks".*

**An organizational culture of sharing.** A culture of sharing is considered an indispensable element for sustaining communities and the systems dedicated to them. Organizational culture produces a system of rules or norms that drive the individuals' behaviours. For example

> *"[…] it is this cultural part that makes more of difficulty, […] to manage to make people enter into a mode of sharing, that it is the most difficult."*

and

> *"[…] we met complex problems, not for technical reasons, but because of the cultural aspects."*

**Support in the use of supporting technologies.** Training and communication for the members of communities in the use of supporting technologies are seen as factors in the success of community-based KMS. Thus:

> *"[…] communication has not been made a priority objective [...] aspects of the management and capitalization of knowledge requires a lot of communication in order to explain to people what it is and what it contains. We must inform people about what exists."*

**Appropriate systems of evaluation and incentives.** Finally, according to those interviewed the evaluation systems and incentive schemes in place can encourage or discourage the sharing via SAGCs. One interviewee said:

> *"If I were assessed and paid on the basis of my contribution to knowledge-sharing and effective feedback, then I'd join the system"*

Another pointed out:

> *"[…] today our managers are not recruited on those criteria […] the way people are selected and sent up the career ladder is at the heart of capitalising on experience".*

Some companies in the sample had introduced incentive schemes to encourage potential SAGC contributors.

**The support of HRM.** According to the people we questioned, the support of the HRM department with regard to the recognition and the processes of evaluation were also essential. For example:

> *"[…] the key is to have a reason to give, for the recognition, for value, because if someone is motivated to do so, then using the tools is not the hardest part"*

## 5 Discussions and Conclusions

From the theoretical viewpoint, the potential contribution of CoPs to KM seems clear (see Table 1 for examples of this). Additionally, the works of Wenger (Wenger 1998b; Wenger 1998a; Wenger 2000; Wenger et al. 2002) readily lend themselves to the identification of various 'factors' that ought to influence the success with which CoPs are formed and the way in which they interact with a 'host' organisation (Bourdon et al. 2003). In contrast to quantitative approaches (Bourdon et al. 2004), the aim of this study was to gain an insight into what practitioners of KM felt were the factors that most influenced the success of CoPs based approaches to KM.

Unsurprisingly, the survey confirmed the importance of CoPs for KM and showed that human factors were an essential component in the development of SAGCs. Of more interest was the emphasis that the interviewees put on the various elements of what we have called the communal model of KM. For example, Wenger (1998b) presented CoPs as containing a number of inherent tensions which he termed dualities. The academic literature on KM and CoPs (Hildreth and Kimble 2002; Kimble and Hildreth 2005) tends to place the participation - reification duality at the centre when dealing with CoPs as a mechanism for Managing Knowledge. The results of the survey however seem to indicate that other aspects, such as those related to organizational structure and, more broadly, 'management' issues, are of more importance.

Similarly, it is interesting to note that, in relation to SAGCs, many of the issues that were raised were 'generic' issues of the type that have seen before in relation to other forms of Information Systems (Kimble and McLoughlin 1995). During the late 1990s and in the very early years of this millennium, there was a tendency to see KMS as some new and exotic form of Information System. Since then it has become clear that KBS face many of the same issues as other forms of IS (e.g. Abdullah et al. 2006) and, in contrast to findings on CoPs and KM, here the survey seems to back the literature.

So, what of the future? The different features of the communal models of KM identified in this research could constitute the elements of a system of evaluation for CoPs in KM. Further quantitative analysis of these characteristics in one or several other communities would be an interesting topic for research. Another area that this work appears to highlight is the tensions between the need to manage KM within existing organizational structures and the need for CoPs to grow and develop in their own way. One of the potentially most interesting aspects of this work is the indication it contains that the role of some of the other dualities identified by Wenger, such as the designed-emergent duality, may also play a role in Knowledge Management.